\documentclass[amsmath,jcp,twocolumn,superscriptaddress]{revtex4-1}
\usepackage{color}
\usepackage{amsmath}
\usepackage{amsfonts}
\usepackage{amssymb}
\usepackage{MnSymbol}
\usepackage{graphicx}
\usepackage[caption=false]{subfig}
\usepackage{tikz}
\usetikzlibrary{arrows}

\usepackage{tikz-feynman}
\tikzfeynmanset{compat=1.0.0}

\begin{document}
\title{
Nonlinear optical spectroscopy of open quantum systems
}
\author{Haoran Sun}
\affiliation{Department of Chemistry \& Biochemistry, University of California San Diego, La Jolla, CA 92093, USA}
\author{Upendra Harbola}
\affiliation{Department of Inorganic and Physical Chemistry, Indian Institute of Science, Bangalore 560012, India}
\author{Shaul Mukamel}
\affiliation{Department of Chemistry, University of California Irvine, Irvine, CA 92697, USA}
\author{Michael Galperin}
\email{mgalperin@tauex.tau.ac.il}
\affiliation{School of Chemistry, Tel Aviv University, Tel Aviv 69978, Israel}
\begin{abstract}
Development of experimental techniques at nanoscale resulted
in ability to perform spectroscopic measurements on single-molecule
current carrying junctions. These experiments are natural meeting point
for research fields of optical spectroscopy and molecular electronics.
We present a pedagogical comparison between perturbation theory 
expansion of standard nonlinear 
optical spectroscopy and (non-self-consistent) perturbative diagrammatic
formulation of the nonequilibrium Green's functions method
(NEGF is widely used in molecular electronics)
indicating their similarities and differences.
Comparing the two approaches we argue that optical spectroscopy 
of open quantum systems
has to be analyzed within the more general Green's function formulation.
\end{abstract}

\maketitle

\section{Introduction}
Interaction of matter with light is on the forefront of research due to its
its ability to answer fundamental questions and its applicational promises.
In particular, spectroscopy is an invaluable characterization tool
in studies of properties of molecules and materials~\cite{ebbesen_hybrid_2016,li_plasmon-enhanced_2017,neubrech_surface-enhanced_2017,zrimsek_single-molecule_2017, zhan_plasmon-enhanced_2018,picque_frequency_2019,baiz_vibrational_2020}.
Also, light-matter interaction allows to control electron and energy transfer,
nuclear dynamics, and chemistry in molecular systems as was shown, 
for example, in recent measurements in systems with strong light-matter interaction. 
The latter opens directions for practical utilization of light as a tool for 
quantum technologies~\cite{kimble_quantum_2008,ritter_elementary_2012,yang_molecular_2022}, 
transport~\cite{orgiu_conductivity_2015,lerario_high-speed_2017,rozenman_long-range_2018,hou_ultralong-range_2020,krainova_polaron_2020,balasubrahmaniyam_enhanced_2023} 
and energy transfer~\cite{akulov_long-distance_2018,xiang_intermolecular_2020} enhancement, 
light engineering~\cite{yariv_coupled-resonator_1999,liu_electromagnetically_2017} , 
and as a catalyst for chemical reactions~\cite{hutchison_modifying_2012,shi_enhanced_2018,munkhbat_suppression_2018}. 
Another novel direction is due to recent experimental developments 
in X-ray spectroscopies. X-ray light is capable to provide spatially localized and 
element specific information extending capabilities of established 
nonlinear spectroscopy techniques~\cite{kraus_ultrafast_2018,rouxel_x-ray_2019}.
Finally, quantum properties of light (e.g. photon entanglement) 
recently attracted attention of researchers as
a tool which allows to circumvent many experimental shortcomings
of spectroscopies relying on classical properties of light~\cite{dorfman_nonlinear_2016,gu_photon_2022}.

Originally, spectroscopy experiments where done on molecules in gas phase and/or
on surfaces. Theoretically, the former situation corresponds to isolated quantum system
(molecule) whose dynamics is described within the time-dependent Schr{\" o}dinger
equation. The latter is formally an open system although the only effect of the bath
(surface) on molecule was related to dissipation. Dynamics of such systems traditionally
is described within the quantum master equation in its simplest, 
Redfield/Lindblad, form and light-matter interaction is taken into account
employing bare perturbation theory (PT) expansions. 
This level of treatment is at the heart of
nonlinear optical spectroscopy theory which is widely and
successfully used for description of experiments and 
perturbation theory expansions define the very classification of
optical processes~\cite{mukamel_principles_1995,harbola_frequency-domain_2014,agarwalla_coherent_2015}.

Recent developments in experimental techniques made it possible
to perform spectroscopic measurements in current carrying single-molecule
junctions. For example, such are bias-induced 
electroluminescence~\cite{imada_real-space_2016,imada_single-molecule_2017,miwa_many-body_2019,chikkaraddy_single-molecule_2023}
and Raman scattering~\cite{ioffe_detection_2008,ward_simultaneous_2008,shamai_spectroscopy_2011,ward_vibrational_2011,jaculbia_single-molecule_2020}.
Molecular cavity spectroscopy is also moving in this direction~\cite{chikkaraddy_single-molecule_2016,benz_single-molecule_2016}.
Open character of these systems plays important role in its responses.
Thus, considerations employing bare perturbation theory expansions
in light-matter interaction and  Markov (weak system-bath coupling) descriptions
of system evolution are not always adequate~\cite{gao_optical_2016,gao_simulation_2016}.
Similarly, widely employed in cavity spectroscopy 
non-Hermitian quantum mechanics considerations 
(i.e. theory taking into account dissipation and ignoring fluctuations)
may lead to qualitative failures~\cite{mukamel_exceptional_2023,seshadri_liouvillian_2024}.

Nonequilibrium Green's functions (NEGF) is the method adequately describing evolution
of an open quantum mechanical system. It is the method of choice in
molecular electronics studies which until recently were mostly focused on
description of electron transport in junctions. 
Theoretical descriptions of spectroscopic measurements in single-molecule junctions
resulted in application of methods from 
nonlinear optical spectroscopy and quantum transport communities, 
and naturally raised question of comparison between them~\cite{galperin_photonics_2017}.
Note that while NEGF studies of optical signals obtained employing 
the bare perturbation theory are available on the literature~\cite{yadalam_spontaneous_2019},
consistent NEGF treatment for optics and transport is new.

In our recent work we showed that shortcomings
of the bare perturbation theory expansion can be corrected within 
the NEGF consideration~\cite{mukamel_flux-conserving_2019}. 
The correction deals with imposing conservation laws 
following celebrated works by Kadanoff and Baym~\cite{baym_conservation_1961,baym_self-consistent_1962}.
The approach implies self-consistent procedure in
treating light-matter interaction which reflects physics
of mutual influence of light and matter and thus requires
significant changes in classification of optical processes
accepted in nonlinear spectroscopy studies.
At the same time, it was noted in a number of
publications that lowest (second order) bare diagrammatic expansion 
does satisfy conservation laws in open systems~\cite{viljas_electron-vibration_2005,de_la_vega_universal_2006,yadalam_energy_2020}.
Bare diagrammatic expansion while not equivalent nevertheless is
close to philosophy of bare perturbation theory consideration
accepted in nonlinear optical spectroscopy.

Here, we discuss general (any order) approach to formulating
conserving bare diagrammatic expansion within NEGF
and compare it with the bare perturbation theory formulation
 standard in spectroscopy studies.
 The comparison allows to extend traditional classification
 of optical processes to the case of open quantum systems.
 Structure of the paper is the following.
 In Section~\ref{model} we introduce model of an open quantum system and
 discuss treatment of its responses within the bare perturbation theory
 and within the NEGF. Numerical results illustrating conserving character 
 of the bare diagrammatic approach are presented in Section~\ref{numres}.
 Conclusions are drawn in Section~\ref{conclude}.


\section{Nonlinear optical spectroscopy}\label{model}
\subsection{Model}

\begin{figure}[b]
\centering\includegraphics[width=\linewidth]{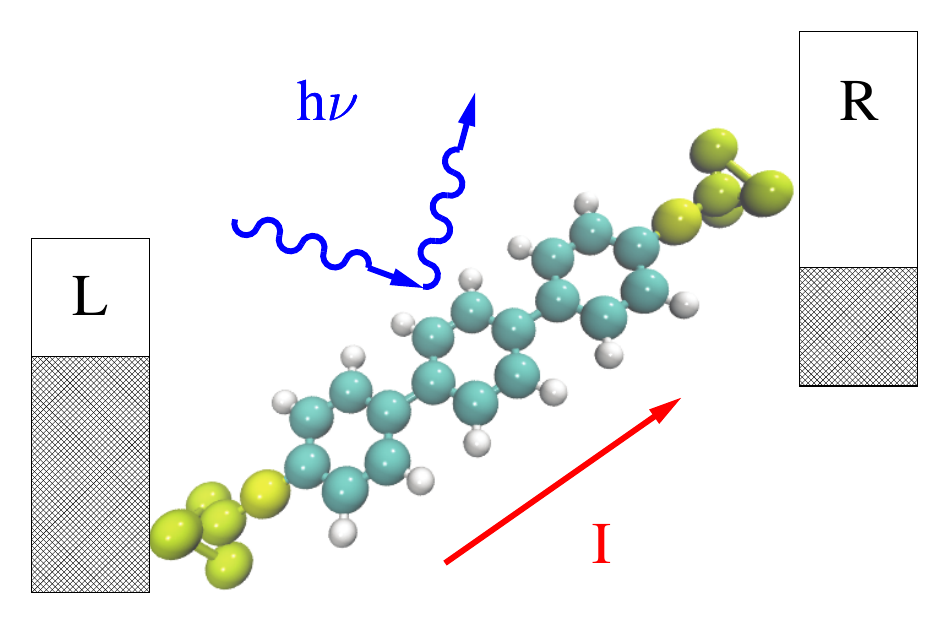}
\caption{\label{fig1}
 Optical spectroscopy of an open quantum system.
}
\end{figure}

We consider open quantum system $S$ driven by external radiation field $rad$
(see Fig.~\ref{fig1}).
Hamiltonian of the model is
\begin{equation}
\label{H}
\hat H=\hat H_S+\hat H_{rad} + \hat V_{S,rad}
\end{equation}
where $\hat H_S$ and $\hat H_{rad}$ are Hamiltnians of the system and 
radiation field, respectively. $\hat V_{S,rad}$ couples between the two
subsystems. 
The system is a junction which consists of molecule $M$ coupled set 
of contacts $K$. Radiation field is modeled as a set of modes $\alpha$.
Explicit expressions are (here and below $\hbar=k_B=1$)
\begin{equation}
\label{H_S_rad}
\begin{split}
\hat H_S &= \sum_{m\in M} \varepsilon_m\hat d_m^\dagger\hat d_m
+\sum_K\sum_{k\in K}\varepsilon_k\hat c_k^\dagger\hat c_k
\\ &
+\sum_{m\in M}\sum_K\sum_{k\in K}\left( V_{mk}\hat d_m^\dagger\hat c_k
   +\mbox{H.c.}\right)
\\
\hat H_{rad} &= \sum_{\alpha\in rad} \omega_\alpha\hat a_\alpha^\dagger\hat a_\alpha
\\
\hat V_{S,rad} &= -\hat P\sum_{\alpha\in rad}
\left(E_\alpha\hat a_\alpha+ \mbox{H.c.}\right)
\end{split}
\end{equation}
Here, $\hat d_m^\dagger$ ($\hat d_m$) and $\hat c_k^\dagger$ ($\hat c_k$)
creates (annihilates) electron in molecular orbital $m$ and contact state $k$, 
respectively.
\begin{equation}
\label{P} 
\hat P \equiv \sum_{m_1,m_2\in M}\mu_{m_1m_2}\hat d_{m_1}^\dagger\hat d_{m_2}
\end{equation}
is the molecular polarization operator with $\mu_{m_1m_2}$ being 
the matrix element of molecular dipole.
$\hat a_\alpha^\dagger$ ($\hat a_\alpha$) creates (annihilates)
excitation in mode $\alpha$ of the field,
and $E_\alpha=i\left(2\pi\hbar\omega_\alpha/V\right)^{1/2}$ is the field amplitude.

We note in passing that non-interacting (molecular orbitals) form of 
the Hamiltonian $\hat H_M$ is chosen for simplicity. Generalization 
to include intra-molecular interactions and/or couplings to Bose
degrees of freedom other than photons
(phonons, electron-hole excitations, thermal baths, etc.)
would require either consideration of additional diagrams
or implementation of many-body flavors of NEGF~\cite{cohen_greens_2020}.

Below we will be interested in simulation of 
electron ($e$) and photon ($p$) particle ($I$) and energy ($J$) fluxes
\begin{equation}
\label{fluxes}
\begin{split}
I_e^K(t) &\equiv -\sum_{k\in K}\frac{d}{dt}\langle \hat c_k^\dagger(t)\hat c_k(t)\rangle
\\
J_e^K(t) &\equiv -\sum_{k\in K}\varepsilon_k\,\frac{d}{dt}\langle \hat c_k^\dagger(t)\hat c_k(t)\rangle
\\
I_p(t) &\equiv -\sum_{\alpha\in rad}\frac{d}{dt}\langle \hat a_\alpha^\dagger(t)\hat a_\alpha(t)\rangle
\\
J_p(t) &\equiv -\sum_{\alpha\in rad}\omega_\alpha\,\frac{d}{dt}\langle \hat a_\alpha^\dagger(t)\hat a_\alpha(t)\rangle
\end{split}
\end{equation}
Here, $\langle\ldots\rangle$ indicates quantum mechanical and statistical average and operators are in the Heisenberg picture.


\subsection{The PT formulation}\label{PT}
Markov evolution and bare perturbative expansion are the two main assumptions
of the traditional theory of nonlinear optical spectroscopy. The former is
caused by similarity between Markov form of the quantum master equation (QME),
\begin{equation}
\label{QME}
\frac{d}{dt}\lvert\rho(t)\rrangle = - i\mathcal{L}(t)\lvert\rho(t)\rrangle,
\end{equation}
and the Schr{\" o}dinger equation which was employed at earlier stages of 
spectroscopy theory. Here, 
\begin{equation}
\begin{split}
\mathcal{L}(t)&=\mathcal{L}_M+\sum_K\left(\mathcal{L}_K+\mathcal{L}_{MK}\right)
+\mathcal{L}_{rad}+\mathcal{L}_{S,rad}
\\ &
\equiv \mathcal{L}_0+\mathcal{L}_{S,rad}
\end{split}
\end{equation}
is Liouvillian which describes evolution of molecular system
(free isolated molecule dynamics $\mathcal{L}_M$) and
radiation field (free radiation field dynamics $\mathcal{L}_{rad}$)
in presence of contacts
(dissipators $\mathcal{L}_K$).
Coupling between molecule and the field is described
by $\mathcal{L}_{S,rad}$.

Markov QME (\ref{QME}) yields simple solution
which in interaction picture is
\begin{equation}
\lvert\rho^I(t)\rrangle=T\,\exp\left[-i\int_{t_0}^{t}dt'\, \mathcal{L}^I_{S,rad}(t')\right]\lvert\rho^I(t_0)\rrangle
\end{equation}
Expansion of evolution operator in Taylor series yields expression
for system density which accounts for light-matter interaction in
order-by-order (bare perturbation theory) form
\begin{equation}
\begin{split}
\lvert\rho^I(t)\rrangle &=\sum_{n=0}^{\infty}(-i)^{n}\int_{t_0}^{t}dt_1\int_{t_0}^{t_1}dt_2\ldots
\int_{t_0}^{t_{n-1}}dt_n\,
\\ &\quad
\mathcal{L}^I_{S,rad}(t_1)\,\mathcal{L}^I_{S,rad}(t_2)\ldots \mathcal{L}^I_{S,rad}(t_n)\,\hat\rho^I(t_0)
\\ &
\equiv \sum_{n=0}^{\infty}\lvert\rho^{I\, (n)}(t)\rrangle
\end{split}
\end{equation}
Using this form in expressions (\ref{P})-(\ref{fluxes}) yields bare perturbation
theory expansions for system properties
\begin{equation}
\label{flux_QME}
\begin{split}
I_e^{K}(t) &= \llangle N_e\rvert i\mathcal{L}_{K}\lvert\rho(t)\rrangle
\equiv \sum_{n=0}^{\infty} I_e^{K\, (n)}(t)
\\
J_e^K(t) &= \llangle H_e\rvert i\mathcal{L}_{K}\lvert\rho(t)\rrangle
\equiv \sum_{n=0}^{\infty} J_e^{K\, (n)}(t)
\\
I_p(t) &= \llangle N_p\rvert i\mathcal{L}_{S,rad}\lvert\rho(t)\rrangle
\equiv \sum_{n=0}^{\infty} I_p^{(n)}(t)
\\
J_p(t) &= \llangle H_p\rvert i\mathcal{L}_{S,rad}\lvert\rho(t)\rrangle
\equiv \sum_{n=0}^{\infty} J_p^{(n)}(t)
\end{split}
\end{equation}
Here, 
$\hat N_e=\sum_{m\in M}\hat d_m^\dagger\hat d_m$  and
 $\hat N_p=\sum_{\alpha\in rad}\hat a_\alpha^\dagger\hat a_\alpha$  is
the electron and photon number operator, respectively. 
$\hat H_e=\sum_{m\in M} \varepsilon_m\hat d_m^\dagger\hat d_m$
and $\hat H_p\equiv\hat H_{rad}=\sum_{\alpha\in rad}\omega_\alpha\hat a_\alpha^\dagger\hat a_\alpha$.
Note opposite sign as compared to traditional definition of optical flux in spectroscopy
studies.
Expansion of $I_p(t)$, Eq.(\ref{flux_QME}), is the basis of description and classification of optical
processes in traditional nonlinear optical spectroscopy.

In many situations radiation field is treated classically.
In this case (\ref{QME}) is equation for molecular density operator only
(this is contrary to total - light and matter - density operator of quantum consideration),
contribution $\mathcal{L}_{S,rad}$ becomes time-dependent,
and term $\mathcal{L}_{rad}$ drops. 


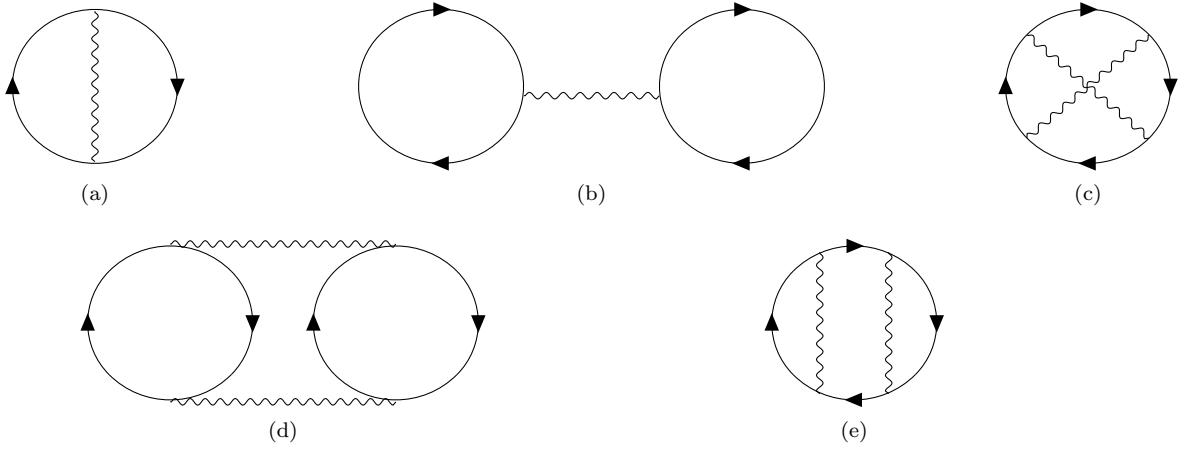
\begin{figure*}[htbp]
{\centering
\hspace*{1cm}
 \subfloat[][]{
\begin{tikzpicture}
\begin{feynman}
\vertex (t);
\vertex [below=3cm of t] (b);
\vertex [below=0.5cm of t] (tb);
\vertex [above=0.5cm of b] (ba);
\vertex [below left=1.1481cm of t] (x1);
\vertex [below right=1.1481cm of t] (x2);
\vertex [above right=1.1481cm of b] (x3);
\vertex [above left=1.1481cm of b] (x4);
\diagram*[layered layout] {
(tb) -- [photon] (ba),
(x1) -- [quarter left] (x2),
(x2) -- [fermion, quarter left] (x3),
(x3) -- [quarter left] (x4),
(x4) -- [fermion, quarter left] (x1),
};
\end{feynman}
\end{tikzpicture}     
}
\hfill
\subfloat[][]{
\begin{tikzpicture}
\begin{feynman}
\vertex (t1);
\vertex [below=3cm of t1] (b1);
\vertex [below=0.45cm of t1] (tb1);
\vertex [above=0.45cm of b1] (ba1);
\vertex [below left=1.1481cm of t1] (x1);
\vertex [below right=1.1481cm of t1] (x2);
\vertex [above right=1.1481cm of b1] (x3);
\vertex [above left=1.1481cm of b1] (x4);
\vertex [below right=1.1481cm of x1] (c1);
\vertex[right=1.1cm of c1] (lr);
\vertex [right=4cm of t1] (t2);
\vertex [below=3cm of t2] (b2);
\vertex [below=0.45cm of t2] (tb2);
\vertex [above=0.45cm of b2] (ba2);
\vertex [below left=1.1481cm of t2] (x5);
\vertex [below right=1.1481cm of t2] (x6);
\vertex [above right=1.1481cm of b2] (x7);
\vertex [above left=1.1481cm of b2] (x8);
\vertex [below right=1.1481cm of x5] (c2);
\vertex[left=1.1cm of c2] (rl);
\diagram*[layered layout] {
(lr) -- [photon] (rl),
(x1) -- [fermion, quarter left] (x2),
(x2) -- [quarter left] (x3),
(x3) -- [fermion, quarter left] (x4),
(x4) -- [quarter left] (x1),
(x5) -- [fermion, quarter left] (x6),
(x6) -- [quarter left] (x7),
(x7) -- [fermion, quarter left] (x8),
(x8) -- [quarter left] (x5),
};
\end{feynman}
\end{tikzpicture}     
}
\hfill
\subfloat[][]{
\begin{tikzpicture}
\begin{feynman}
\vertex (t);
\vertex [below=3cm of t] (b);
\vertex [below left=1.1481cm of t] (x1);
\vertex [below right=1.1481cm of t] (x2);
\vertex [above right=1.1481cm of b] (x3);
\vertex [above left=1.1481cm of b] (x4);
\diagram*[layered layout] {
(x1) -- [photon] (x3),
(x2) -- [photon] (x4),
(x1) -- [fermion, quarter left] (x2),
(x2) -- [fermion, quarter left] (x3),
(x3) -- [fermion, quarter left] (x4),
(x4) -- [fermion, quarter left] (x1),
};
\end{feynman}
\end{tikzpicture}             
}
\hspace*{1cm}
}\\
{\centering
\hspace*{2cm}
\subfloat[][]{
\begin{tikzpicture}
\begin{feynman}
\vertex (t1);
\vertex [below=3cm of t1] (b1);
\vertex [below=0.45cm of t1] (tb1);
\vertex [above=0.45cm of b1] (ba1);
\vertex [below left=1.1481cm of t1] (x1);
\vertex [below right=1.1481cm of t1] (x2);
\vertex [above right=1.1481cm of b1] (x3);
\vertex [above left=1.1481cm of b1] (x4);
\vertex [right=3cm of t1] (t2);
\vertex [below=3cm of t2] (b2);
\vertex [below=0.45cm of t2] (tb2);
\vertex [above=0.45cm of b2] (ba2);
\vertex [below left=1.1481cm of t2] (x5);
\vertex [below right=1.1481cm of t2] (x6);
\vertex [above right=1.1481cm of b2] (x7);
\vertex [above left=1.1481cm of b2] (x8);
\diagram*[layered layout] {
(tb1) -- [photon] (tb2),
(x1) -- [quarter left] (x2),
(x2) -- [fermion, quarter left] (x3),
(x3) -- [quarter left] (x4),
(x4) -- [fermion, quarter left] (x1),
(ba1) -- [photon] (ba2),
(x5) -- [quarter left] (x6),
(x6) -- [fermion, quarter left] (x7),
(x7) -- [quarter left] (x8),
(x8) -- [fermion, quarter left] (x5),
};
\end{feynman}
\end{tikzpicture}     
}
\hfill
\subfloat[][]{
\begin{tikzpicture}
\begin{feynman}
\vertex (t);
\vertex [below=3cm of t] (b);
\vertex [below left=1.1481cm of t] (x1);
\vertex [below right=1.1481cm of t] (x2);
\vertex [above right=1.1481cm of b] (x3);
\vertex [above left=1.1481cm of b] (x4);
\vertex [above right=0.5cm of x1] (x1ar);
\vertex [below right=0.5cm of x4] (x4br);
\vertex [above left=0.5cm of x2] (x2al);
\vertex [below left=0.5cm of x3] (x3bl);
\vertex [below=0.1cm of x1ar] (x11);
\vertex [above=0.1cm of x4br] (x44);
\vertex [below=0.1cm of x2al] (x22);
\vertex [above=0.1cm of x3bl] (x33);
\diagram*[layered layout] {
(x11) -- [photon] (x44),
(x22) -- [photon] (x33),
(x1) -- [fermion, quarter left] (x2),
(x2) -- [fermion, quarter left] (x3),
(x3) -- [fermion, quarter left] (x4),
(x4) -- [fermion, quarter left] (x1),
};
\end{feynman}
\end{tikzpicture}     
}
\hspace*{4cm}
}
\caption{$\Phi$ functional for electron-photon interaction. 
Shown are (a)-(b) second and (c)-(d) fourth order contributions to $\Phi^{ep}$.
Panel (e) shows additional fourth order contribution to $\Phi^{ep\, (r)}$.
Directed solid line represents electron propagator $G$ or $G^{(0)}$.
Wavy line represents photon propagator $F^{(0)}$, both directions are possible 
for each wavy line.}\label{figPhi}
\end{figure*}


\subsection{The NEGF formulations}\label{NEGF}
Central object in NEGF formulation is single-particle Green's function
of the open quantum system
\begin{equation}
\label{G}
 G_{m_1m_2}(\tau_1,\tau_2)=-\langle T_c\, \hat d_{m_1}(\tau_1)\,\hat d_{m_2}^\dagger(\tau_2)\rangle
\end{equation}
where $T_c$ is the Keldysh contour ordering operator
and $\tau_{1,2}$ are the contour variables.
Its dynamics is described by the Dyson equation
which can be written in two equivalent forms~\cite{stefanucci_nonequilibrium_2013} 
\begin{widetext}
\begin{align}
\label{Dyson}
G_{m_1m_2}(\tau_1,\tau_2) &= G^{(0)}_{m_1m_2}(\tau_1,\tau_2)
+\sum_{m_3,m_4}\int_c d\tau_3\int_c d\tau_4\, G^{(0)}_{m_1m_3}(\tau_1,\tau_3)\,
\Sigma^{ep}_{m_3m_4}(\tau_3,\tau_4)\, G_{m_4m_2}(\tau_4,\tau_2)
\\ 
\label{Dyson_r}
&= G^{(0)}_{m_1m_2}(\tau_1,\tau_2)
+\sum_{m_3,m_4}\int_c d\tau_3\int_c d\tau_4\, G^{(0)}_{m_1m_3}(\tau_1,\tau_3)\,
\Sigma^{ep\,(r)}_{m_3m_4}(\tau_3,\tau_4)\, G^{(0)}_{m_4m_2}(\tau_4,\tau_2)
\end{align}
\end{widetext}
Here, $G^{(0)}$ is Green's function of the junction in absence of radiation field,
$\Sigma^{ep}$ and $\Sigma^{ep\,(r)}$ are proper and reducible forms of the self-energy
which describes electron-photon interaction. Due to many-body character 
of the interaction, exact expression for the self-energy is not available.
Diagrammatic technique is a way to obtain approximate (up to a particular order
in the interaction) solutions. For the approximation to be conserving
the self-energy should be $\Phi$-derivable~\cite{stefanucci_nonequilibrium_2013} 
\begin{align}
\label{Sep}
\Sigma^{ep}_{m_1m_2}(\tau_1,\tau_2) &= -\frac{\delta\Phi^{ep}}{\delta G_{m_2m_1}(\tau_2,\tau_1)}
\\
\label{Sep_r}
\Sigma^{ep\,(r)}_{m_1m_2}(\tau_1,\tau_2) &= -\frac{\delta\Phi^{ep\,(r)}}{\delta G^{(0)}_{m_2m_1}(\tau_2,\tau_1)}
\end{align}
Here, $\Phi^{ep}$ is the sum of all connected and topologically inequivalent
skeleton diagrams, while $\Phi^{ep\,(r)}$ is the sum
of all connected and topologically inequivalent vacuum diagrams
(no dressing). Diagrams up to fourth order in electron-photon interaction 
are shown in Fig.~\ref{figPhi}. Explicit expressions for the self-energies are given
in Appendix~\ref{appSE} and corresponding diagrams are shown in
Fig.~\ref{figSE}. Note that $\Phi$ functional presented in Fig.~\ref{figPhi}b
does not contribute to self-energy because zero-frequency photons do not exist.
However, the functional does contribute to kernel expression
(see explanation below); the contribution is presented in Fig.~\ref{figK}b.   

To calculate energy fluxes, Eqs.~(\ref{fluxes}), one needs to know also
two-particle Green's function
\begin{equation}
\label{G2}
\begin{split}
& G_{m_1m_2,m_3m_4}(\tau_1,\tau_2;\tau_3,\tau_4) \equiv
\\ &\qquad
-\langle T_c\, \hat d_{m_1}(\tau_1)\,\hat d_{m_2}(\tau_2)\,
\hat d_{m_4}^\dagger(\tau_4)\,\hat d_{m_3}^\dagger(\tau_3) \rangle
\end{split}
\end{equation}
Its dynamics is described by 
(we use shorthand notation with $1$ standing for $m_1$ and $\tau_1$, etc.)
\begin{widetext}
\begin{align}
\label{BS}
G(1,2;3,4) &= G(1,3)G(2,4)-G(1,4)G(2,3)
\\ &
+\int d1'\int d2'\int d3'\int d4'\,
G(1,1')G(3',3)K^{ep}(1',2';3',4')\left[G(2,2')G(4',4)-G(2,4';4,2')\right]
\nonumber \\ 
\label{BS_rr}
&= G(1,3)G(2,4)-G(1,4)G(2,3)
\\ &
+\int d1'\int d2'\int d3'\int d4'\,
G^{(0)}(1,1')G^{(0)}(3',3)K^{ep\, (r)}(1',2';3',4')G^{(0)}(2,2')G^{(0)}(4',4)
\nonumber
\end{align}
\end{widetext}
Here, $K^{ep}$ and $K^{ep\,(r)}$ are irreducible and reducible (in two-particle sense) 
forms of the kernel.

Similar to self-energy, for the approximation to be conserving
the kernel should be $\Phi$-derivable (second derivative)~\cite{stefanucci_nonequilibrium_2013} 
\begin{equation}
\label{Kep}
K^{ep}(1,2;3,4) = \frac{\delta^2\Phi^{ep}}{\delta G(4,2)\,\delta G(3,1)}
\equiv -\frac{\delta\Sigma^{ep}(1,3)}{\delta G(4,2)}
\end{equation}
Reducible kernel, $K^{ep\, (r)}$, is related to the irreducible one, $K^{ep}$,
via the Bethe-Salpeter equation
\begin{align}
\label{Kep_rr}
& K^{ep\,(r)}(1,2;3,4) =  K^{ep}(1,2;3,4)
\nonumber \\ &
-\int d1'\int d2'\int d3'\int d4'\, K^{ep}(1,2';3,4')\, 
\\ &\qquad\,\,\,\times
G(4',1')\, G(3',2')\,
K^{ep\, (r)}(1',2;3',4)
\nonumber
\end{align}
Diagrams up to fourth order in electron-photon interaction 
and explicit expressions for the irreducible kernel are given
in Appendix~\ref{appK} and corresponding diagrams are shown in
Fig.~\ref{figK}.

The knowledge of Green's functions (\ref{Dyson}) and (\ref{BS})
or (\ref{Dyson_r}) and (\ref{BS_rr})
allows us to calculate 
molecular polarization, and particle and energy fluxes
\begin{widetext}
\begin{align}
\label{flux_NEGF} 
I_e^K(t) &= 2\,\mbox{Re}\sum_{m,m'\in M}\int_{t_0}^t dt'\,\big(
\Sigma_{mm'}^{K\, <}(t,t')\, G_{m'm}^{>}(t',t) -
\Sigma_{mm'}^{K\, >}(t,t')\, G_{m'm}^{<}(t',t) 
\big)
\nonumber \\ 
J_e^K(t) &= 2\,\mbox{Re}\sum_{k\in K}\varepsilon_k\sum_{m,n\in M}\int_{t_0}^t dt'\,
V_{mk}V_{km'}
\big( g_k^{<}(t,t')\, G_{m'm}^{>}(t',t) - g_k^{>}(t,t')\, G_{m'm}^{<}(t',t) \big)
\nonumber \\
I_p(t) &= -2\,\mbox{Im}
\sum_{m,n,m',n'\in M}\int_{t_0}^t dt'\,
\big(
\Pi_{mn,m'n'}^{<}(t,t')\, G_{mn',nm'}^{-+-+}(t,t';t,t')-
\Pi_{mn,m'n'}^{>}(t,t')\, G_{mn',nm'}^{+-+-}(t,t',t,t')
\big)
\\ &
\equiv -2\,\mbox{Re}\sum_{\alpha\in rad}\int_{t_0}^t dt'\, \lvert E_\alpha\rvert^2\big(
F_{\alpha}^{(0)\, <}(t,t')\, G_{PP}^{>}(t't)-
F_{\alpha}^{(0)\, >}(t,t')\, G_{PP}^{<}(t't)
\big)
\nonumber \\
J_p(t) &= -2\,\mbox{Im}\sum_{\alpha\in rad}\omega_\alpha\sum_{m,n,m',n'\in M}\int_{t_0}^t dt'\,\mu_{mn}^{*}\mu_{m'n'}\lvert E_\alpha\rvert^2\left(
F_{\alpha}^{(0)\, <}(t,t')\, G_{mn',nm'}^{-+-+}(t,t';t,t')-
F_{\alpha}^{(0)\, >}(t,t')\, G_{mn',nm'}^{+-+-}(t,t';t,t')
\right)
\nonumber 
\end{align}
\end{widetext}
Here, $G^{>/<}$ is the lesser/greater projection of the Green's function
(\ref{Dyson}) or (\ref{Dyson_r}), $\Sigma^K$ and $g_k$
are, respectively, the self-energy due to coupling to contact $K$ 
and Green's function of free electron in state $k$ in the contact
\begin{equation}
\begin{split}
g_k(\tau_1,\tau_2) &\equiv -i\langle T_c\,\hat c_k(\tau_1)\,\hat c_k^\dagger(\tau_2)\rangle_0
\\
\Sigma^{K}_{m_1m_2}(\tau_1,\tau_2) &\equiv \sum_{k\in K}
V_{m_1k}g_k(\tau_1,\tau_2)V_{km_2}
\end{split}
\end{equation}
$G_{mn',m'n}$ are the $-++-$ and $+--+$ projections
($-$/$+$indicates forward/backward branch of the Keldysh contour)
of the two-particle Green's function
and $\Pi$ and $F_\alpha^{(0)}$ are the photon-induced interaction
and Green's function of free photon defined in Eqs.~(\ref{Pi}) and (\ref{F}),
respectively. $G_{PP}$ is correlation function of the molecular polarization operator
\begin{equation}
G_{PP}(\tau_1,\tau_2)\equiv -i\langle T_c\,\hat P(\tau)\,\hat P^\dagger(\tau')\rangle
\end{equation}
The following observations are notable:
\begin{enumerate}
\item Because electron-photon is the only many-body interaction of 
the model (\ref{H})-(\ref{H_S_rad}) we discuss only its contributions
to approximate expressions for self-energy and kernel.
In presence of additional many-body interactions 
(e.g., electron-electron or electron-phonon) additional contributions
to these expressions should be taken into account. Moreover,
interactions resulting in time-nonlocal correlations in electron
subsystem may lead to contributions mixed with those of the radiation
field. This would affect discussion of optical processes presented below.
\item Formulation (\ref{Dyson}) and (\ref{BS})
is self-consistent procedure. For example, single-particle Green's function 
(\ref{Dyson}) depends on self-energy, and
self-energy, Eqs.~(\ref{Sep2}) and (\ref{Sep4}), depends on Green's function.
This self-consistency accounts for mutual influence of electrons and radiation
field in determining state of the system. The formulation is standard in
studies of quantum transport and we used it in our previous work
to discuss nonlinear spectroscopy in open systems~\cite{mukamel_flux-conserving_2019}.
\item Formulation (\ref{Dyson_r}) and (\ref{BS_rr})
treats interaction with radiation field order-by-order. 
Thus, results for molecular polarization and fluxes
are obtained as sum of contributions of different orders in
the light-matter interaction. In this sense, (\ref{Dyson_r}) and (\ref{BS_rr}) 
is closer in spirit to scattering type consideration of traditional nonlinear 
spectroscopy. 
\item For the non-interacting molecular model represented by Hamiltonian 
$\hat H_S$, Eq.(\ref{H_S_rad}), consideration of the two-particle Green's function
(\ref{G2}) and higher correlations is only relevant in the self-consistent
formulation (\ref{Dyson}) and (\ref{BS}). Single-particle
Green's function (\ref{G}) is the only correlation function required in  
the formulation (\ref{Dyson_r}) and (\ref{BS_rr}), because all
higher order correlation functions can be expressed in terms of
single-particle Green's function employing the Wick's theorem.
\item As we discussed in Refs.~\cite{gao_optical_2016,mukamel_flux-conserving_2019},
employing bare perturbation expansion, which is standard in nonlinear 
spectroscopy studies, in the formulation (\ref{Dyson}) and (\ref{BS})
leads to violation of conservation laws.
Below we use formulation  (\ref{Dyson_r}) and (\ref{BS_rr})
to compare NEGF with the bare perturbation theory treatment of traditional 
nonlinear optical spectroscopy.
\end{enumerate}


\subsection{Comparison between PT and NEGF}\label{compare}
Here, we follow classification of optical processes introduced in 
Chapter~9 of Ref.~\cite{mukamel_principles_1995}.
We note in passing that classification of second order process
as absorption and fourth order as spontaneous emission
used in Ref.~\cite{mukamel_principles_1995}
is based on assumptions of occupied ground and unoccupied
excited molecular states.
For nonequilibrium with bias induced occupation of excited state and/or
depletion of ground state the classification does not hold anymore.
Nevertheless, below we use traditional names for the second and
fourth order optical processes.

\subsubsection{Absorption of a quantum field}
In the PT formulation, linear absorption is given by $n=1$ 
contribution to photon flux in Eq.(\ref{flux_QME})
\begin{align}
\label{abs_Ip_QME}
I_p^{abs}(t)&=\llangle N_p\rvert \int_{t_0}^{t}dt_1\,
\mathcal{L}_{S,rad}^I(t)\,\mathcal{L}_{S,rad}^I(t_1)\lvert\rho(t_0)\rrangle
\\ &=
\int_{t_0}^{t}dt_1\,\mbox{Tr}\big\{\left[\left[\hat N_p,\hat H_{S,rad}^I(t)\right],\hat H_{S,rad}^I(t_1)\right]\hat\rho(t_0)\big\}
\nonumber
\end{align}
Here, $\hat\rho$ is the total density matrix of the system.
Note that initially molecule and radiation field are assumed to be decoupled
\begin{equation}
\hat\rho(t_0)=\hat\rho_M(t_0)\otimes\hat\rho_{rad}(t_0)
\end{equation}
Note also that usually absorption is considered driven by a particular 
mode $\alpha$ of radiation field (only one mode of radiation field is included in the Liouvillian/Hamiltonian).

Within the (\ref{Dyson_r}) and (\ref{BS_rr}) NEGF formulation,
absorption is given by the second order (in light-matter coupling)
contribution to photon flux. It is obtained from (\ref{flux_NEGF}) by 
substituting 
\begin{align}
G_{PP}^{(0)}(\tau,\tau') &\equiv 
-i\sum_{m_1,m_2,m_3,m_4\in M}\mu_{m_1m_2}\,\mu_{m_3m_4}^{*}\,
\\ &\qquad\times
G_{m_3m_1}^{(0)}(\tau',\tau)\, G_{m_2m_4}^{(0)}(\tau,\tau')
\nonumber
\end{align}
in place of $G_{PP}$ in the expression. This leads to
\begin{align}
\label{abs_Ip_NEGF}
&I_p^{(2)}(t) =-2\,\mbox{Re}\sum_{\alpha\in rad} \int_{t_0}^t dt'\, \lvert E_\alpha\rvert^2
\\ &\times
\big(
F_{\alpha}^{(0)\, <}(t,t')\, G^{(0)\, >}_{PP}(t',t)-F_{\alpha}^{(0)\, >}(t,t')\, G_{PP}^{(0)\, <}(t',t)
\big)
\nonumber
\end{align}

It is straightforward to show that after evaluating commutators in 
the PT result, Eq.(\ref{abs_Ip_QME}),
and separating molecular and radiation field degrees of freedom one gets 
exactly structure of the NEGF expression, Eq.(\ref{abs_Ip_NEGF}).
The only difference is lack of hybridization information in the PT correlation
function of molecular polarization (often this information is added  
to the correlation function in an approximate {\em ad hoc} manner), 
while the NEGF expression accounts for the hybridization exactly.

\subsubsection{Spontaneous light emission spectroscopy}
In the PT formulation, SLE is given by $n=3$ contribution to photon flux 
in Eq.(\ref{flux_QME})
\begin{align}
\label{SLE_Ip_QME}
&I_p^{SLE}(t)=\llangle N_p\rvert \int_{t_0}^{t}dt_1\int_{t_0}^{t_1}dt_2\int_{t_0}^{t_2}dt_3\,
\\ &\qquad
\mathcal{L}_{S,rad}^I(t)\,\mathcal{L}_{S,rad}^I(t_1)\,\mathcal{L}_{S,rad}^I(t_2)\,\mathcal{L}_{S,rad}^I(t_3)\,\lvert\rho(t_0)\rrangle
\nonumber \\ &=
\int_{t_0}^{t}dt_1\int_{t_0}^{t_1}dt_2\int_{t_0}^{t_2}dt_3\,
\mbox{Tr}\big\{\big[\big[\big[\big[\hat N_p,\hat H_{S,rad}^I(t)\big],
\nonumber \\ &\qquad\qquad
\hat H_{S,rad}^I(t_1)\big],\hat H_{S,rad}^I(t_2)\big],\hat H_{S,rad}^I(t_3)\big]\hat\rho(t_0)\big\}
\nonumber
\end{align}

Within the (\ref{Dyson_r}) and (\ref{BS_rr}) NEGF formulation,
SLE is given by the fourth order (in light-matter coupling)
contribution to photon flux. It is obtained from (\ref{flux_NEGF}) by 
substituting
\begin{align}
 &G_{PP}^{(2)}(\tau,\tau') \equiv
 -i\sum_{m_1,m_2,m_3,m_4\in M}\mu_{m_1m_2}\mu^{*}_{m_3m_4}\bigg(
\nonumber \\ &
 G^{(0)}_{m_3m_1}(\tau',\tau)\, G^{(2)}_{m_2m_4}(\tau,\tau') +
 G^{(2)}_{m_3m_1}(\tau',\tau)\, G^{(0)}_{m_2m_4}(\tau,\tau')
\nonumber  \\ &
 -\sum_{n_1,n_2,n_3,n_4}\int_c d\tau_1\int_c d\tau_2\int_c d\tau_3\int_c d\tau_4\, 
 \\ &\qquad
 G^{(0)}_{m_2n_1}(\tau,\tau_1)\, G^{(0)}_{n_3m_1}(\tau_3,\tau)\,
 K^{ep\, (r)\, (2)}_{n_1n_2;n_3n_4}(\tau_1,\tau_2;\tau_3,\tau_4)\,
\nonumber  \\ &\qquad
G^{(0)}_{m_3n_2}(\tau',\tau_2)\, G^{(0)}_{n_4m_4}(\tau_4,\tau')
\bigg)
\nonumber
\end{align}
in place of $G_{PP}$ in the expression. 
Here, 
\begin{align}
& G^{(2)}_{m_1m_2}(\tau_1,\tau_2)= \sum_{m_3,m_4} \int_c d\tau_3\int_c d\tau_4\, 
 \\ &\qquad\qquad\qquad
G^{(0)}_{m_1m_3}(\tau_1,\tau_3)\, \Sigma^{ep\, (r)\, (2)}_{m_3m_4}(\tau_3,\tau_4)\, G^{(0)}_{m_4m_2}(\tau_4,\tau_2)
\nonumber
\end{align}
and expression for $K^{ep\, (r)\, (2)}$ is given in Eq.(\ref{Kep2}).
This leads to
\begin{align}
\label{SLE_Ip_NEGF}
&I_p^{(4)}(t) =-2\,\mbox{Re}\sum_{\alpha\in rad} \int_{t_0}^t dt'\, \lvert E_\alpha\rvert^2
\\ &\times
\big(
F_{\alpha}^{(0)\, <}(t,t')\, G^{(2)\, >}_{PP}(t',t)-F_{\alpha}^{(0)\, >}(t,t')\, G_{PP}^{(2)\, <}(t',t)
\big)
\nonumber
\end{align}

Direct comparison between the two approaches is less straightforward in this case
because multi-time electron correlation function is evaluated within different
schemes: the quantum regression theorem (QRT) in the QME case
vs. the Bethe-Salpeter equation in the NEGF formulation. Nevertheless, the following
conclusions are possible
\begin{enumerate}
\item Standard nonlinear spectroscopy as well as  (\ref{Dyson_r}) 
and (\ref{BS_rr}) NEGF formulation
are similar because both are based on PT expansion of multi-time
correlation function of molecular polarization.
Their main difference is in the way to evaluate the multi-time correlation function.
\item By the very construction, QRT
evaluates multi-time correlation function as a sequence of blocks
between adjacent times when molecule interacts with radiation field. 
Electronic correlations due to coupling to Fermi baths for pair of times
in different blocks are missing in such scheme.
They are preserved in the NEGF formulation.
\item In particular, inability by QRT to describe correlation across the blocks
implies that radiation field induced interactions 
(such as, e.g., presented by diagram in Fig.~\ref{figSE}c) 
cannot be accounted properly in the standard nonlinear spectroscopy formulation.
We note in passing that limitations of the QRT when applied to
description of open quantum systems were extensively discussed
in the literature~\cite{talkner_failure_1986,guarnieri_quantum_2014,jin_non-markovian_2016,cosacchi_accuracy_2021}.
\item Another point of critical difference between the PT and NEGF
is related to accounting for all possible dicouplings within PT,
while only connected diagrams contribute within the NEGF.
Thus, among other processes the PT accounts for virtual 
excitation in the radiation field, which neither affect the system, 
nor related to its physics.  
\item Finally, we note the difference in language of standard nonlinear
spectroscopy PT formulation and that of diagrammatic NEGF expansion.
Famous double sided Feynman diagrams of the PT formulation
are related to particular arrangements of times when molecule interacts 
with radiation field on the Keldysh contour. Within the NEGF,
such arrangements are called projections while word diagrams 
is used to indicate different physical processes (see Fig.~\ref{figSE}). 
\end{enumerate}


\section{Numerical results}\label{numres}

In Ref.~\cite{gao_optical_2016} we demonstrated violation of conservation 
laws in an open system
when bare perturbation theory is employed in (\ref{Dyson}) and (\ref{BS}) formulation
of the NEGF. Here, we illustrate that perturbative diagrammatic expansion
(\ref{Dyson_r}) and (\ref{BS_rr}) is conserving. 

\begin{figure}[t]
\centering\includegraphics[width=\linewidth]{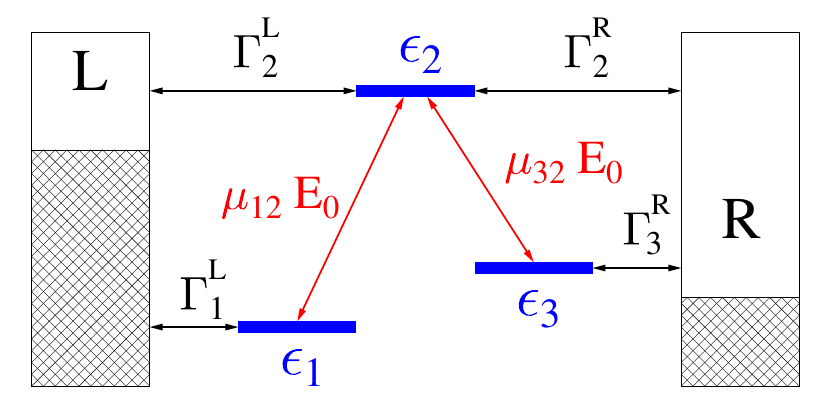}
\caption{\label{fig_model}
(Color online) Model of an open donor-bridge-acceptor system
with light assisted tunneling.
}
\end{figure}

We employ the same donor-bridge-acceptor molecular model
as in Ref.~\cite{gao_optical_2016} (see Fig.~\ref{fig_model}).
Parameters of the simulation are 
temperature ($k_BT=0.25$), 
positions of molecular levels 
($\varepsilon_1=-5$, $\varepsilon_2=5$, $\varepsilon_3=-2$).
Electron escape rates into metallic contacts are taken within the
wide-band approximation
($\Gamma_1^L=\Gamma_3^R=1$, $\Gamma_2^L=\Gamma_2^R=0.1$).
Energy escape rate into radiation bath is modeled as
\begin{equation}
 \gamma(\omega)=\gamma_0\left(\frac{\omega}{\omega_0}\right)^2
 e^{(\omega_0^2-\omega^2)/\omega_C^2}
\end{equation}
with $\gamma_0=0.01$ and $\omega_C=10$.
Molecule is pumped by a monochromatic laser 
characterized by its frequency tuned to resonance with molecular 
transition between levels $2$ and $3$, 
$\omega_0=\varepsilon_2-\varepsilon_3=7$,
intensity $N_0=1$, and bandwidth $\delta=0.1$.
Fermi energy is taken as origin, $E_F=0$, and bias is assumed to be applied symmetrically, $\mu_{L,R}=E_F\pm\lvert e\rvert V_{sd}/2$.
Simulations are performed on energy grid spanning region from $-100$ to $100$
with step $0.01$.
Here, energy is given in terms of an arbitrary unit of energy $E_0$.
Results for particle and energy fluxes are presented in terms of 
flux units $I_0=1/t_0$ and $J_0=E_0/t_0$, respectively, where 
$t_0=\hbar/E_0$ is unit of time.

\begin{figure}[b]
\centering\includegraphics[width=\linewidth]{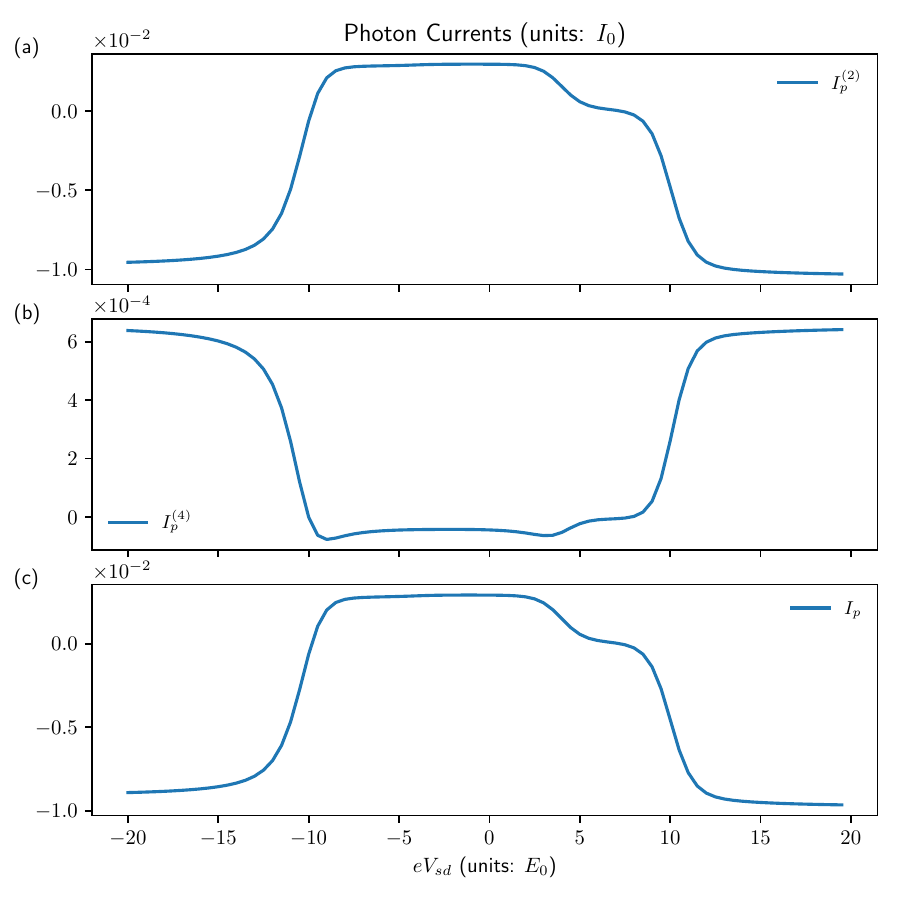}
\caption{\label{fig_photon}
(Color online) Photon flux in DBA junction of Fig.~\ref{fig_model} 
vs applied bias.
Shown are (a) second and (b) fourth contributions.
Panel (c) shows total photon flux.
}
\end{figure}

\begin{figure}[t]
\centering\includegraphics[width=\linewidth]{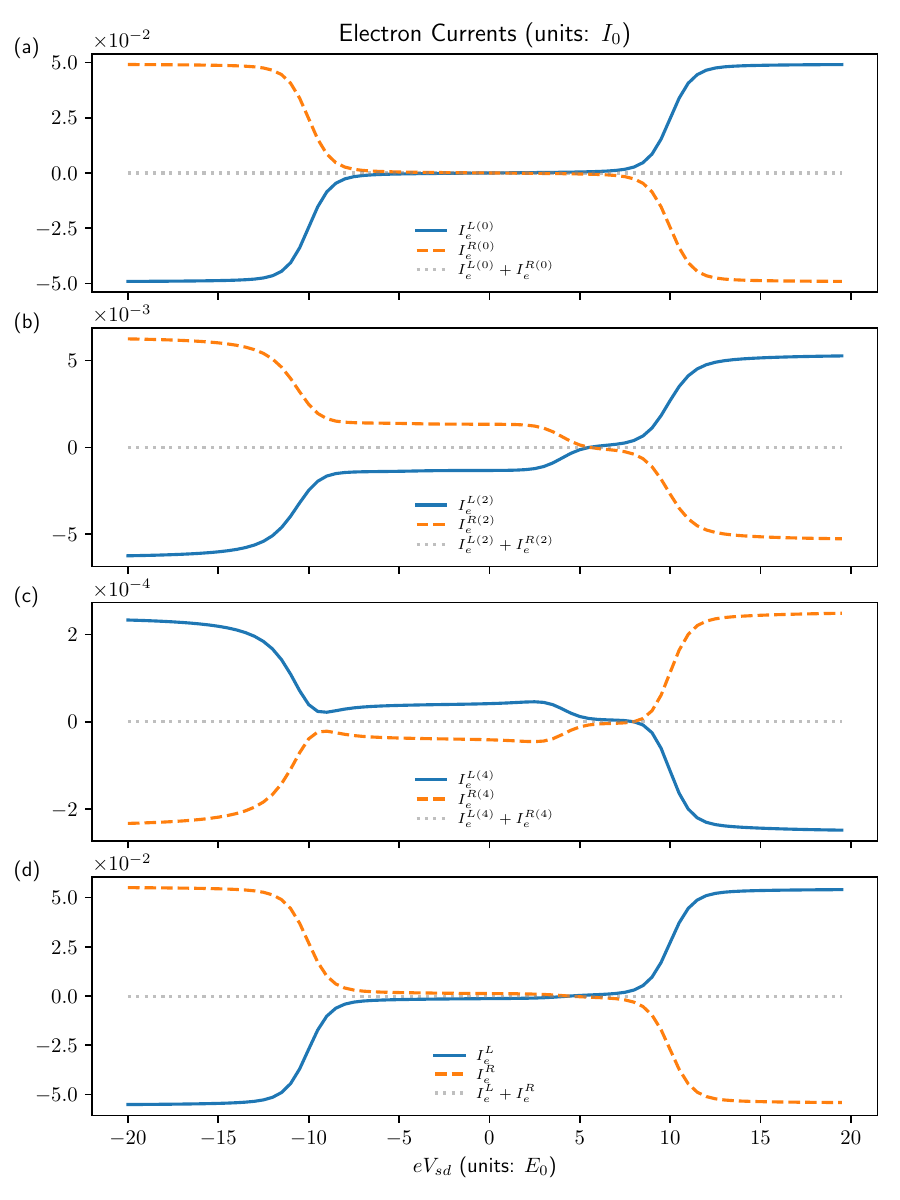}
\caption{\label{fig_particle}
(Color online) Electron (particle) currents in DBA junction of Fig.~\ref{fig_model} 
vs applied bias.
Shown are currents at left $I_e^L$ (solid line, blue) and right $I_e^R$ (dashed line, red) interfaces for (a) zero, (b) second, and (c) fourth orders in light-matter interaction. 
Panel (d) shows total electron currents. 
In each panel, sum of the left and right currents
is indicated by dotted line (grey).
}
\end{figure}

Figure~\ref{fig_photon} shows photon flux vs. applied bias. 
As mentioned earlier second order optical process (panel a) is pure absorption
and fourth order optical process (panel b) is pure emission
only at and near equilibrium. Significant changes in the total photon flux (panel c)
occur when chemical potentials cross molecular resonances.

Figure~\ref{fig_particle} illustrates charge conservation. At steady-state 
amount of charge entering the system at left interface should be compensated by 
amount of charge leaving the system at right interface. That is, $I_e^L=-I_e^R$.
As expected, currents at the two interfaces (solid and dashed lines) sum
to zero (dotted line). Note that conserving character of the approximation
holds at every order of perturbation theory (see panels a, b, and c).

\begin{figure}[t]
\centering\includegraphics[width=\linewidth]{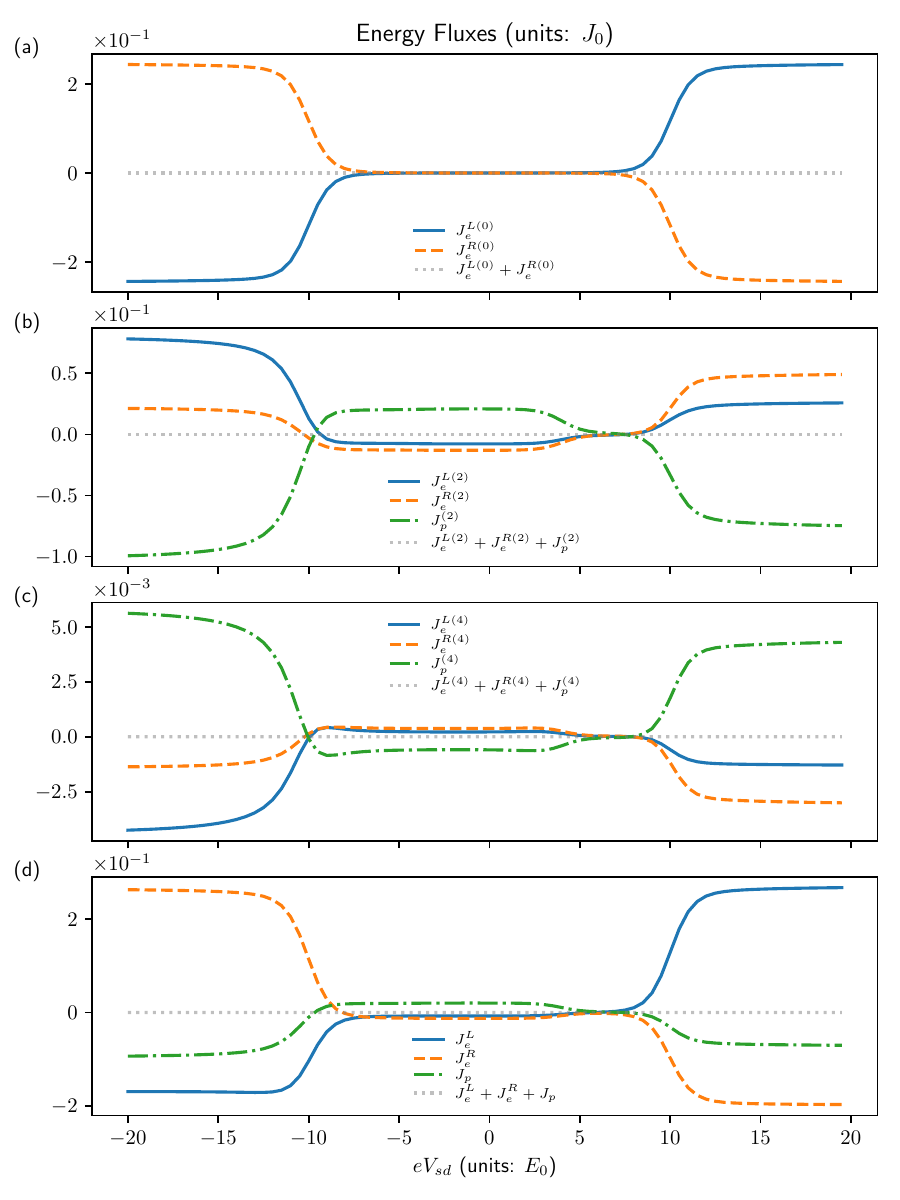}
\caption{\label{fig_energy}
(Color online) Energy currents in DBA junction of Fig.~\ref{fig_model} vs applied bias.
Shown are electron energy currents at left $J_e^L$ (solid line, blue) and right $J_e^R$ (dashed line, red) interfaces, and photon energy current $J_p$
(dash-dotted line, green) for (a) zero, (b) second, and (c) fourth orders
in light-matter interaction. 
Panel (d) shows corresponding total energy currents.
In each panel, sum of the energy currents is indicated by dotted line (grey).
}
\end{figure}

Figure~\ref{fig_energy} illustrates energy conservation. At steady-state 
amount of energy entering the system should be compensated by 
amount of charge leaving the system. Energy enters (leaves) the system
from (to) left and right contacts and from (to) radiation field modes.  
 That is, $J_e^L+J_e^R+J_p=0$.
As expected, energy currents at the two interfaces (solid and dashed lines) 
plus energy current from radiation modes (dash-dotted line) sum
to zero (dotted line). Again, conservation holds separately for
contributions of every order in light-matter interaction.
 
\section{Conclusions}\label{conclude}
Development of experimental techniques at nanoscale resulted in
combination of research fields of optical spectroscopy and molecular electronics.
Today, theoretical studies of spectroscopy in open quantum systems
use methods traditionally coming from nonlinear optical spectroscopy 
and quantum transport. The former employ Markov QME description 
of molecular dynamics and are based on bare perturbation theory 
expansion of light-matter interaction. Molecular electronics,
which historically developed from quantum transport measurements,
employs diagrammatic expansion of the NEGF method.
Usual NEGF formulation is based on consideration of dressed 
diagrams which corresponds to accounting for mutual influence of 
light and matter degrees of freedom on their dynamics.
In our previous studies we noted that employing bare perturbation 
expansion in such formulation leads to violation of conservation laws.

Here, we consider an alternative - bare diagrams formulation of the NEGF.
This description is close in spirit to the bare PT of traditional nonlinear spectroscopy
and thus is advantageous in attempt to explore connections between
theoretical methods of the two communities.
Comparing the two methods we find close similarity in overall structure of 
the approaches: both are based on bare perturbation theory expansion
in light-matter interaction, in each order of the expansion both require
evaluation of a multi-time correlation function of molecular polarization operators.
Main difference between the methods is in the way the latter evaluation is 
performed: Markov QME description of traditional nonlinear optical
spectroscopy relies on the quantum regression theorem (QRT), while
NEGF employs Wick's theorem and diagrammatic methods.

We note that limitation of the QRT in description of dynamics in open
quantum systems was widely discussed in the literature.
In particular, we argue that QRT describes dynamics as sequence of 
propagations between times when molecule interacts with the radiation field
and as such is not adequate in description of coherences in molecular system
across the times. For example, it cannot properly describe two-time
correlation function with one time before and the other after the time of
interaction. Note that such correlation functions naturally appear in
non-Markov NEGF description. Note also that in case of quantum 
light the limitation affects description of radiation field induced intra-molecular
interactions (see, e.g., diagram in Fig.~\ref{figSE}c).

On formal level we note difference in terminology between
the PT and NEGF treatments. In particular, traditional nonlinear optical
spectroscopy employs double sided Feynman diagrams for description
and classification of optical processes. These diagrams are particular 
arrangements of light-matter interaction times on the Keldysh contour.
Within the NEGF, such arrangements are called projections,
while word diagram is reserved for description of different physical 
processes in light-matter interaction.

We conclude that traditional way of processes classification
accepted in nonlinear optical spectroscopy is preserved 
when describing open quantum systems.  
However, it is important to employ non-Markov description of 
dynamics in such systems. Technically, this results in 
using diagrammatic methods of the NEGF in evaluation 
of multi-time correlation functions instead of the 
QRT  as is employed in traditional spectroscopy studies.

\begin{acknowledgments}
This material is based upon work supported by the National Science Foundation
  under Grant No. CHE-2154323.
\end{acknowledgments}

\appendix
\begin{widetext}
\section{Self-energies due to electron-photon interaction}\label{appSE}

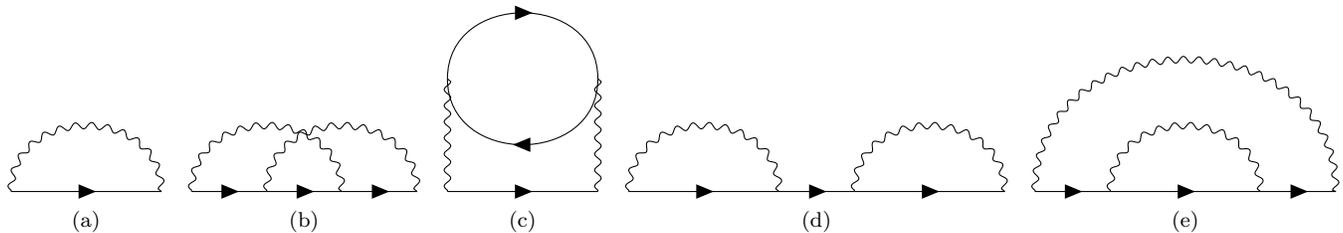
\begin{figure}[htbp]
{\centering
 \subfloat[][]{
\begin{tikzpicture}
\begin{feynman}
\vertex (x1);
\vertex [right=2cm of x1] (x2);
\diagram*[layered layout] {
(x1) -- [fermion] (x2),
(x1) -- [photon, half left] (x2),
};
\end{feynman}
\end{tikzpicture}     
}
\hfill
\subfloat[][]{
\begin{tikzpicture}
\begin{feynman}
\vertex (x1);
\vertex [right=1cm of x1] (x2);
\vertex [right=1cm of x2] (x3);
\vertex [right=1cm of x3] (x4);
\diagram*[layered layout] {
(x1) -- [fermion] (x2) -- [fermion] (x3) -- [fermion] (x4),
(x1) -- [photon, half left] (x3),
(x2) -- [photon, half left] (x4),
};
\end{feynman}
\end{tikzpicture}             
}
\hfill
\subfloat[][]{
\begin{tikzpicture}
\begin{feynman}
\vertex (x1);
\vertex [right=2cm of x1] (x2);
\vertex [above=1.5cm of x1] (x3);
\vertex [above=1.5cm of x2] (x4);
\diagram*[layered layout] {
(x1) -- [fermion] (x2),
(x3) -- [fermion, half left] (x4),
(x4) -- [fermion, half left] (x3),
(x1) -- [photon] (x3),
(x2) -- [photon] (x4),
};
\end{feynman}
\end{tikzpicture}     
}
\hfill
\subfloat[][]{
\begin{tikzpicture}
\begin{feynman}
\vertex (x1);
\vertex [right=2cm of x1] (x2);
\vertex [right=1cm of x2] (x3);
\vertex [right=2cm of x3] (x4);
\diagram*[layered layout] {
(x1) -- [fermion] (x2) -- [fermion] (x3) -- [fermion] (x4),
(x1) -- [photon, half left] (x2),
(x3) -- [photon, half left] (x4),
};
\end{feynman}
\end{tikzpicture}     
}
\hfill
\subfloat[][]{
\begin{tikzpicture}
\begin{feynman}
\vertex (x1);
\vertex [right=1cm of x1] (x2);
\vertex [right=2cm of x2] (x3);
\vertex [right=1cm of x3] (x4);
\diagram*[layered layout] {
(x1) -- [fermion] (x2) -- [fermion] (x3) -- [fermion] (x4),
(x1) -- [photon, half left] (x4),
(x2) -- [photon, half left] (x3),
};
\end{feynman}
\end{tikzpicture}     
}
}
\caption{Self-energy due to electron-photon interaction. 
Shown are second (panel a) and fourth (panels b and c) order contributions to $\Sigma^{ep}$.
Panels d and e show additional fourth order contributions to $\Sigma^{ep\, (r)}$.}\label{figSE}
\end{figure}

Second order expressions are
\begin{align}
\label{Sep2}
\Sigma^{ep\,(2)}_{m_1m_2}(\tau_1,\tau_2) &=
i\sum_{n_1,n_2\in M} G_{n_1n_2}(\tau_1,\tau_2)
\left[\Pi_{n_1m_1,n_2m_2}(\tau_1,\tau_2)+\Pi_{m_2n_2,m_1n_1}(\tau_2,\tau_1)\right]
\\
\label{Sep_r2}
\Sigma^{ep\, (r)\,(2)}_{m_1m_2}(\tau_1,\tau_2) &=
i\sum_{n_1,n_2\in M} G^{(0)}_{n_1n_2}(\tau_1,\tau_2)
\left[\Pi_{n_1m_1,n_2m_2}(\tau_1,\tau_2)+\Pi_{m_2n_2,m_1n_1}(\tau_2,\tau_1)\right]
\end{align}
Fourth order expressions
\begin{align}
\label{Sep4}
\Sigma^{ep\,(4)}_{m_1m_2}(\tau_1,\tau_2) &=
\sum_{n_1,n_2,n_3,n_4,m_3,m_4}\int_c d\tau_3\int_c d\tau_4
\\ &
\bigg(
G_{n_1n_2}(\tau_1,\tau_2)\, G_{m_3m_4}(\tau_3,\tau_4)\, G_{n_4n_3}(\tau_4,\tau_3)
\nonumber \\ &\qquad\times
\left[\Pi_{n_1m_1,n_3m_3}(\tau_1,\tau_3)+\Pi_{m_3n_3,m_1n_1}(\tau_3,\tau_1)\right]
\left[\Pi_{m_2n_2,m_4n_4}(\tau_2,\tau_4)+\Pi_{n_4m_4,n_2m_2}(\tau_4,\tau_2)\right]
\nonumber \\ &
-G_{n_1m_3}(\tau_1,\tau_3)\, G_{n_3n_4}(\tau_3,\tau_4)\, G_{m_4n_2}(\tau_4,\tau_2)
\nonumber \\ &\qquad\times
\left[\Pi_{n_1m_1,n_4m_4}(\tau_1,\tau_4)+\Pi_{m_4n_4,m_1n_1}(\tau_4,\tau_1)\right]
\left[\Pi_{n_3m_3,n_2m_2}(\tau_3,\tau_2)+\Pi_{m_2n_2,m_3n_3}(\tau_2,\tau_3)\right]
\bigg)
\nonumber \\
\label{Sep_r4}
\Sigma^{ep\,(r)\,(4)}_{m_1m_2}(\tau_1,\tau_2) &=
\sum_{n_1,n_2,n_3,n_4,m_3,m_4}\int_c d\tau_3\int_c d\tau_4
\\ &
\bigg(
G^{(0)}_{n_1n_2}(\tau_1,\tau_2)\, G^{(0)}_{m_3m_4}(\tau_3,\tau_4)\, G^{(0)}_{n_4n_3}(\tau_4,\tau_3)
\nonumber \\ &\qquad\times\quad
\left[\Pi_{n_1m_1,n_3m_3}(\tau_1,\tau_3)+\Pi_{m_3n_3,m_1n_1}(\tau_3,\tau_1)\right]
\left[\Pi_{m_2n_2,m_4n_4}(\tau_2,\tau_4)+\Pi_{n_4m_4,n_2m_2}(\tau_4,\tau_2)\right]
\nonumber \\ &
-G^{(0)}_{n_1m_3}(\tau_1,\tau_3)\, G^{(0)}_{n_3n_4}(\tau_3,\tau_4)\, G^{(0)}_{m_4n_2}(\tau_4,\tau_2)
\nonumber \\ &\qquad\times
\bigg\{
\left[\Pi_{n_1m_1,n_4m_4}(\tau_1,\tau_4)+\Pi_{m_4n_4,m_1n_1}(\tau_4,\tau_1)\right]
\left[\Pi_{n_3m_3,n_2m_2}(\tau_3,\tau_2)+\Pi_{m_2n_2,m_3n_3}(\tau_2,\tau_3)\right]
\nonumber \\ &\qquad\quad
+
\left[\Pi_{n_1m_1,m_3n_3}(\tau_1,\tau_3)+\Pi_{n_3m_3,m_1n_1}(\tau_3,\tau_1)\right]
\left[\Pi_{m_4n_4,n_2m_2}(\tau_4,\tau_2)+\Pi_{m_2n_2,n_4m_4}(\tau_2,\tau_4)\right]
\nonumber \\ &\qquad\quad
+
\left[\Pi_{n_1m_1,n_2m_2}(\tau_1,\tau_2)+\Pi_{m_2n_2,m_1n_1}(\tau_2,\tau_1)\right]
\left[\Pi_{n_3m_3,n_4m_4}(\tau_3,\tau_4)+\Pi_{m_4n_4,m_3n_3}(\tau_4,\tau_3)\right]
\bigg\}
\bigg)
\nonumber 
\end{align}
Here,
\begin{equation}
\label{Pi}
\Pi_{m_1n_1,m_2n_2}(\tau_1,\tau_2) = \mu_{m_1n_1}^{*}\mu_{m_2n_2}
\sum_{\alpha\in rad}\lvert E_\alpha\rvert^2 F_\alpha^{(0)}(\tau_1,\tau_2)
\end{equation}
is photon-induced interaction between electronic transitions $m_1n_1$
and $m_2n_2$, and
\begin{equation}
\label{F}
F_\alpha^{(0)}(\tau_1,\tau_2) = -i\langle T_c\, \hat a_\alpha(\tau_1)\,\hat a_\alpha^\dagger(\tau_2)\rangle_0
\end{equation}
is Green's function of free photon of mode $\alpha$.


\section{Irreducible kernel due to electron-photon interaction}\label{appK}

\begin{figure}[htbp]
{\centering
 \subfloat[][]{
\begin{tikzpicture}
\begin{feynman}
\vertex (x1);
\vertex [below=2cm of x1] (x2);
\diagram*[layered layout] {
(x1) -- [photon] (x2),
};
\end{feynman}
\end{tikzpicture}     
}
\hfill
 \subfloat[][]{
\begin{tikzpicture}
\begin{feynman}
\vertex (x1);
\vertex [above=1cm of x1] (x2);
\vertex [right=1.5cm of x2] (x3);
\diagram*[layered layout] {
(x1) -- [fermion, opacity=0] (x2),
(x2) -- [photon] (x3),
};
\end{feynman}
\end{tikzpicture}     
}
\hfill
\subfloat[][]{
\begin{tikzpicture}
\begin{feynman}
\vertex (x1);
\vertex [below=0.5cm of x1] (x2);
\vertex [below=1cm of x2] (x3);
\vertex [below=0.5cm of x3] (x4);
\diagram*[layered layout] {
(x1) -- [photon] (x2),
(x2) -- [fermion, half left] (x3),
(x3) -- [fermion, half left] (x2),
(x3) -- [photon] (x4),
};
\end{feynman}
\end{tikzpicture}             
}
\hfill
\subfloat[][]{
\begin{tikzpicture}
\begin{feynman}
\vertex (x1);
\vertex [right=1.5cm of x1] (x4);
\vertex [below=2cm of x1] (x3);
\vertex [right=1.5cm of x3] (x2);
\diagram*[layered layout] {
(x1) -- [photon] (x4),
(x3) -- [photon] (x2),
(x1) -- [fermion] (x3),
(x2) -- [fermion] (x4),
};
\end{feynman}
\end{tikzpicture}             
}
\hfill
\subfloat[][]{
\begin{tikzpicture}
\begin{feynman}
\vertex (x1);
\vertex [right=1.5cm of x1] (x4);
\vertex [below=2cm of x1] (x3);
\vertex [right=1.5cm of x3] (x2);
\diagram*[layered layout] {
(x1) -- [photon] (x2),
(x3) -- [photon] (x4),
(x1) -- [fermion] (x3),
(x2) -- [fermion] (x4),
};
\end{feynman}
\end{tikzpicture}             
}
\hfill
\subfloat[][]{
\begin{tikzpicture}
\begin{feynman}
\vertex (x1);
\vertex [below=1cm of x1] (x2);
\vertex [below left=1.41cm of x2] (x3);
\vertex [below right=1.41cm of x2] (x4);
\diagram*[layered layout] {
(x1) -- [photon] (x2),
(x4) -- [fermion] (x2),
(x2) -- [fermion] (x3),
(x3) -- [photon] (x4),
};
\end{feynman}
\end{tikzpicture}     
}
\hfill
\subfloat[][]{
\begin{tikzpicture}
\begin{feynman}
\vertex (x1);
\vertex [above=1cm of x1] (x2);
\vertex [above left=1.41cm of x2] (x3);
\vertex [above right=1.41cm of x2] (x4);
\diagram*[layered layout] {
(x1) -- [photon] (x2),
(x3) -- [fermion] (x2),
(x2) -- [fermion] (x4),
(x3) -- [photon] (x4),
};
\end{feynman}
\end{tikzpicture}     
}
\hfill
\subfloat[][]{
\begin{tikzpicture}
\begin{feynman}
\vertex (x1);
\vertex [right=1.5cm of x1] (x4);
\vertex [below=2cm of x1] (x3);
\vertex [right=1.5cm of x3] (x2);
\diagram*[layered layout] {
(x1) -- [fermion] (x4),
(x2) -- [fermion] (x3),
(x1) -- [photon] (x2),
(x3) -- [photon] (x4),
};
\end{feynman}
\end{tikzpicture}     
}
}
\caption{Irreducible kernel due to electron-photon interaction. 
Shown are second (panels a-b) and fourth (panels b-h) order contributions to 
$K^{ep}$.}
\label{figK}
\end{figure}
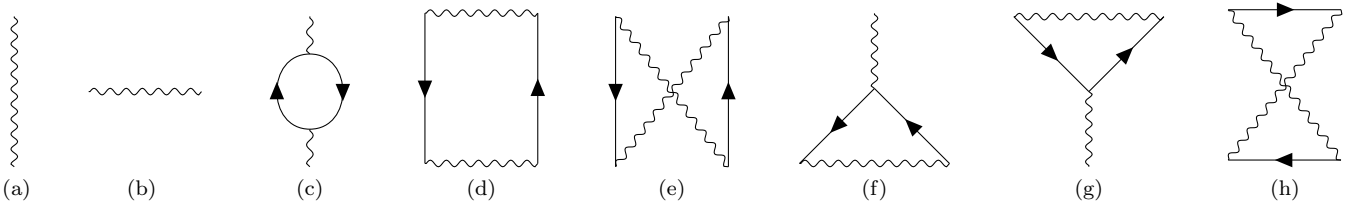

Second order expression is
\begin{align}
\label{Kep2}
K^{ep\, (2)}_{m_1m_2;m_3m_4}(\tau_1,\tau_2;\tau_3,\tau_4) =&
-i\delta(\tau_1,\tau_4)\delta(\tau_2,\tau_3)
\left[\Pi_{m_4m_1,m_2m_3}(\tau_1,\tau_2)+\Pi_{m_3m_2,m_1m_4}(\tau_2,\tau_1)\right]
\\ &
+i\delta(\tau_1,\tau_3)\delta(\tau_2,\tau_4)
\left[\Pi_{m_3m_1,m_2m_4}(\tau_1,\tau_2)+\Pi_{m_4m_2,m_1m_3}(\tau_2,\tau_1)\right]
\nonumber
\end{align}
Fourth order expression is
\begin{align}
& K^{ep\, (4)}_{m_1m_2;m_3m_4}(\tau_1,\tau_2;\tau_3,\tau_4) =
\\ &
-\delta(\tau_1,\tau_4)\delta(\tau_2,\tau_3)\sum_{n_5,n_6,m_5,m_6}
\int_c d\tau_5\int_cd\tau_6\, G_{n_5n_6}(\tau_5,\tau_6)G_{m_6m_5}(\tau_6,\tau_5)
\nonumber \\ &\qquad\qquad\qquad\qquad\qquad\times
\left[\Pi_{m_4m_1,m_5n_5}(\tau_1,\tau_5)+\Pi_{n_5m_5,m_1m_4}(\tau_5,\tau_1)\right]
\left[\Pi_{m_3m_2,n_6m_6}(\tau_2,\tau_6)+\Pi_{m_6n_6,m_2m_3}(\tau_6,\tau_2)\right]
\nonumber \\ &-
\sum_{n_1,n_2,n_3,n_4} G_{n_1n_3}(\tau_1,\tau_3) G_{n_2n_4}(\tau_2,\tau_4)
\nonumber \\ &\qquad\qquad\qquad\qquad\qquad\times
\bigg(
\left[\Pi_{n_1m_1,n_4m_4}(\tau_1,\tau_4)+\Pi_{m_4n_4,m_1n_1}(\tau_4,\tau_1)\right]
\left[\Pi_{m_3n_3,m_2n_2}(\tau_3,\tau_2)+\Pi_{n_2m_2,n_3m_3}(\tau_2,\tau_3)\right]
\nonumber \\ &\qquad\qquad\qquad\qquad\qquad\,\,\,\,\, +
\left[\Pi_{n_1m_1,m_2n_2}(\tau_1,\tau_2)+\Pi_{n_2m_2,m_1n_1}(\tau_2,\tau_1)\right]
\left[\Pi_{m_3n_3,n_4m_4}(\tau_3,\tau_4)+\Pi_{m_4n_4,n_3m_3}(\tau_4,\tau_3)\right]
\bigg)
\nonumber \\ &+\delta(\tau_1,\tau_4)\sum_{m,n,n_2,n_3}\int_c d\tau\,
G_{n_2m}(\tau_2,\tau)G_{n,n_3}(\tau,\tau_3)
\nonumber \\ &\qquad\qquad\qquad\qquad\qquad\times
\left[\Pi_{m_4m_1,mn}(\tau_1,\tau)+\Pi_{nm,m_1m_4}(\tau,\tau_1)\right]
\left[\Pi_{n_2m_2,n_3m_3}(\tau_2,\tau_3)+\Pi_{m_3n_2,m_2n_2}(\tau_3,\tau_2)\right]
\nonumber \\ &+\delta(\tau_2,\tau_3)\sum_{m,n,n_1,n_4}\int_c d\tau\,
G_{n_1n}(\tau_1,\tau) G_{mn_4}(\tau,\tau_4)
\nonumber \\ &\qquad\qquad\qquad\qquad\qquad\times
\left[\Pi_{n_1m_1,n_4m_4}(\tau_1,\tau_4)+\Pi_{m_4n_4,m_1n_1}(\tau_4,\tau_1)\right]
\left[\Pi_{mn,m_2m_3}(\tau,\tau_3)+\Pi_{m_3m_2,nm}(\tau_3,\tau)\right]
\nonumber \\ &+\sum_{n_1,n_2,n_3,n_4} G_{n_1n_4}(\tau_1,\tau_4) G_{n_2n_3}(\tau_2,\tau_3)
\nonumber \\ &\qquad\qquad\qquad\qquad\qquad\times
\left[\Pi_{n_1m_1,m_2n_2}(\tau_1,\tau_2)+\Pi_{n_2m_2,m_1n_1}(\tau_2,\tau_1)\right]
\left[\Pi_{m_4n_4,n_3m_3}(\tau_4,\tau_3)+\Pi_{m_3n_3,n_4m_4}(\tau_3,\tau_4)\right]
\nonumber 
\end{align}
\end{widetext}


%

\end{document}